\documentstyle[preprint,epsf]{jpsj}
\def\runtitle{Magnetization plateau in the frustrated spin ladder}
 
\title{Magnetization Plateau in the Frustrated Spin Ladder}
\author{Nobuhisa {\sc Okazaki}, Junji {\sc Miyoshi}$^1$ 
and T\^oru {\sc Sakai}}
\inst{Faculty of Science, Himeji Institute of Technology,\\Kamigouri-cho,
Akou-gun, Hyogo 678-1297, Japan \\ 
Faculty of Science, Okayama University, Okayama 700, Japan$^1$}
\recdate{\today}
\abst{
The magnetization process of the $S$=1/2 antiferromagnetic spin
ladder at $T$=0 is
studied by the exact diagonalization of finite clusters and size scaling
analyses. It is found that a magnetization plateau appears at half the
saturation value ($m=1/2$) in the presence of a sufficiently
large next-nearest-neighbor exchange interaction to yield the
frustration, when the rung coupling is larger than the leg one. 
The phase diagram at $m=1/2$ is given by the analysis based on the 
conformal invariance. 
The magnetization curves are also presented
in several cases.}
\kword{spin ladder, magnetization plateau, frustration}
\begin{document}
\sloppy
\maketitle

The spin ladder has attracted a lot of interest in the fields of the
magnetism and high temperature superconductivity.
Many studies have revealed that the low-temperature properties of a
system are characterized by the spin gap\cite{rf:1,rf:2,rf:3} 
due to large quantum fluctuations.
Recently several theoretical approaches \cite{rf:4,rf:5}
suggested the possibility of
another spin gap induced by the external magnetic field
generally in one-dimensional antiferromagnets.
The spin gap is then observed as a plateau in the magnetization curve.
The condition of the appearance of the plateau 
was rigorously given\cite{rf:4} as 
\begin{eqnarray}
Q(S-m) = {\rm integer},
\label{quantization}
\end{eqnarray}
where $Q$ is the spatial period of the ground state measured by the unit
cell. 
$S$ and $m$ are the total spin and the magnetization per unit cell, 
respectively. 
In fact it was theoretically revealed that the plateau could appear 
in some generalized spin ladders; three-leg\cite{rf:6}, 
dimerized\cite{rf:7} and zigzag\cite{rf:5,rf:8} 
ladders. 
The last one is also equivalent to the bond-alternating chain with 
the next-nearest neighbor exchange interaction which is one of the realistic models of CuGeO$_3$.
We consider another scenario of the plateau in the spin ladder, 
which is due to the symmetric next-nearest neighbor exchange interaction. 
This implies that both diagonal exchanges are equivalently taken into 
account at every plaquette of the ladder. 
In contrast to the zigzag ladder, 
this model has reflection symmetry along the rung,   
realized in the standard spin ladder. 
A strong coupling approach \cite{rf:9}
showed that the plateau also appears
in the spin ladder with the symmetric next-nearest neighbor exchange
interaction, in the large rung coupling limit.
In this paper,
the magnetization process of the $S=1/2$ frustrated antiferromagnetizatic
spin ladder is studied quantitatively 
using the exact diagonalization of finite clusters and size scaling
techniques, to verify the realization of the plateau and 
give the phase diagram at $m=1/2$. 

We consider the $S$=1/2 antiferromagnetic spin ladder with the 
next-nearest-neighbor exchange interaction in a magnetic field
described by the Heisenberg Hamiltonian
\begin{eqnarray}
{\hat H}&=&{\hat H}_0+{\hat H}_Z\\
{\hat H}_0&=&J_1\sum_i^L({\bf S}_{1,i} \cdot {\bf S}_{1,i+1}
+{\bf S}_{2,i} \cdot {\bf S}_{2,i+1})\nonumber\\
& &+J_{\perp}\sum_i^L({\bf S}_{1,i} \cdot {\bf S}_{2,i})\nonumber\\
& &+\alpha J_1\sum_i^L({\bf S}_{1,i}
\cdot {\bf S}_{2,i+1}+{\bf S}_{2,i} \cdot {\bf
S}_{1,i+1})\\
{\hat H}_Z&=&-H\sum_i^L({S_{1,i}^z}+{S_{2,i}^z}),
\end{eqnarray}
under the periodic boundary condition,
where $J_{1}$, $J_{\perp}$ and $\alpha J_1$ denotes the coupling 
constants of the leg, rung and next-nearest neighbor (diagonal) 
exchange interactions, respectively.(Fig. 1)
\begin{figure}
\begin{center}
\leavevmode
\epsfxsize=7.0cm
\epsfysize=3.0cm
\epsfbox{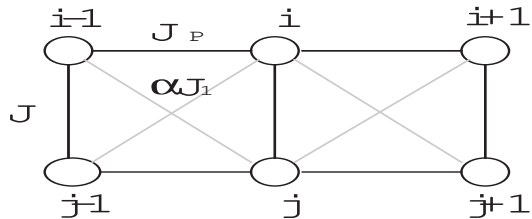}
\caption{Spin ladder with next-nearest neighbor exchange interactions
along the diagonals
}
\label{fig.1}
\end{center}
\end{figure}
We put $J_{\perp}$=1 in the following.
${\cal H}_Z$ is the Zeeman term where $H$ denotes the magnetic field
along the $z$-axis and the energy is specified by the eigenvalue $M$
of the conserved quantity $\sum_{i}{({S_{1,i}^z}+
{S_{2,i}^z)}}$.
We calculate, by the Lanczos algorithm, the lowest eigenvalue $E(L,M)$ of the
Hamiltonian ${\hat H}_0$ in the subspace of constant
$M$=$\sum_{i}{({S_{1,i}^z}+
{S_{2,i}^z)}}$.
($L$ is even and $L \leq 12$.)
The macroscopic magnetization is represented by $m=M/L$.
The magnetic excitation gap giving $\delta M =\pm1$ for the total 
Hamiltonian {\cal H} is given by 
\begin{eqnarray}
\Delta_{\pm}=E(L,M{\pm}1)-E(L,M){\mp}H.
\end{eqnarray}
The width of the plateau in the magnetization curve given by
\begin{eqnarray} 
\Delta&{\equiv}&{\Delta_+}+{\Delta_-} \nonumber \\
      &=&E(L,M+1)+E(L,M-1)-2E(L,M), 
\end{eqnarray} 
is a useful order parameter.\cite{rf:10}
When the system has a plateau, $\Delta$ should converge to a finite 
value $\Delta(\infty)({\neq}0)$ in the infinite length limit 
and the scaled gap $L\Delta$ diverges in proportion to $L$. 
In contrast, 
when the system has no plateau, 
$\Delta$ should decay as $1/L(L{\rightarrow}{\infty})$ and 
the scaled gap $L\Delta$ is independent of $L$.

First, we investigate the possibility of the appearance of the magnetization 
plateau at $m{\equiv}M/L=1/2$.
The scaled gap $L \Delta$ of finite systems ($L=8{\sim}12$) 
with fixed $J_1$ (=0.4) is plotted
against $\alpha$ in Fig. 2(a).
\begin{figure}
\begin{center}
\leavevmode
\epsfxsize=8.0cm
\epsfbox{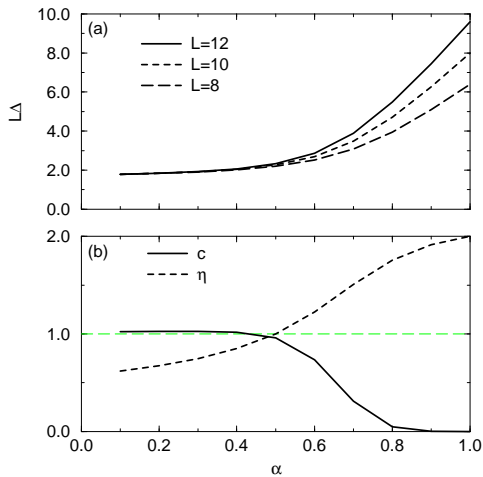}
\caption{(a)Scaled gap $L\Delta$ vs $\alpha$ (b)Conformal anomaly $c$
and exponent $\eta$ vs $\alpha$($J_1=0.4$)}
\label{fig.1}
\end{center}
\end{figure}
In Fig. 2(a), it is apparent that $L\Delta$ increases with increasing
$L$ for $\alpha> 0.5$, while it is almost independent of $L$
around the region $\alpha< 0.5$.
This implies that the plateau would appear for large $\alpha$,
while the system is still gapless for small $\alpha$.
The phase boundary is expected to be characterized by
the Kosterlitz-Thouless (KT) transition. \cite{rf:11}
Let us introduce the critical exponent $\eta$ defined by the spin 
correlation function 
$\langle S^+_0S^-_r \rangle \sim (-1)^r r^{-\eta}$.
Next we estimate the conformal anomaly $c$ in the conformal field theory
(CFT)\cite{rf:12}.
According to the CFT, if the system is massless at $m$, 
the size dependence of the energy has 
the asymptotic form 
\begin{eqnarray}
\frac{1}{L}E(L,M){\sim}{\epsilon}(m)- \frac{\pi}{6}c{\upsilon}_s
\frac{1}{L^2} \hspace{1cm} (L\rightarrow \infty),
\end{eqnarray}
where ${\epsilon}(m)$ is the energy of the bulk system, 
$c$ is the conformal anomaly,
and $\upsilon_s$ is the sound velocity. 
${\upsilon}_s$ is estimated by
\begin{eqnarray}
{\upsilon}_s=\frac{L}{2\pi}[E_{k_1}(L,M)-E(L,M)]+O \left( \frac{1}{L^2}
\right),
\end{eqnarray}
Where $E_{k_1}(L,M)$ is the lowest energy with the momentum 
$k_1\equiv 2\pi /L$. 
CFT predicts the relation 
$\Delta \sim 2\pi \upsilon_s \eta /L \quad (L\rightarrow \infty)$, 
the exponent $\eta$ can be estimated by the 
form \cite{rf:13}
\begin{eqnarray}
\eta=\frac{\Delta}{E_{k_1}(L,M)-E(L,M)}+O \left( \frac{1}{L^2} \right).
\end{eqnarray}
It is expected that 
the conformal anomaly holds $c=1$ in the gapless region.
In Fig. 2(b), $c$ begins to deviate from the value of 1 around
$\alpha{\sim}0.5$, indicating the appearance of a plateau there.
Around the
same $\alpha$ value, $\eta=1$ happens to occur.
An analysis based on the two-dimensional sine-Gordon model\cite{rf:14}
suggested that the exponent $\eta$ should be $1/n^2$ ($n$ is an integer)
at the KT phase boundary.
In ref.14, the isotropic point of the $S=1/2$ $XXZ$ chain was also
mentioned as a good example of the KT transition with $\eta =1$.
We can therefore use the condition $\eta=1$ to determine the phase
boundary below.
\begin{figure}
\begin{center}
\leavevmode
\epsfxsize=7.0cm
\epsfbox{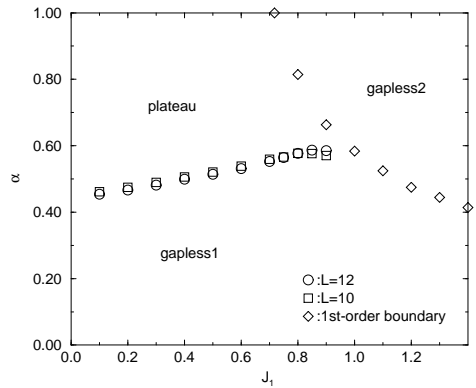}
\caption{
Phase diagram in the $\alpha$-$J_1$ plane at $m=1/2$
}
\label{fig.2}
\end{center}
\end{figure}
The boundaries obtained for $L=10$ and 12 are shown
in Fig. 3.
The $L$ dependence is so small that the calculated boundary is
believed to be very close to the bulk one.
Besides the KT line, we also found the first-order phase boundary, which
was
detected as the energy level crossing in the ground state at $m=1/2$.
Since it is almost independent of $L$, we show the boundary in Fig. 3
only for $L=12$.
The scaled gap analysis similar to Fig. 2(a) revealed that another
gapless phase (gapless2) with $c=1$ exists above the boundary of
the plateau for larger $\alpha$.
Thus the phase diagram for $m=1/2$ consists of the two gapless phases
and
the plateau phase.
The first order line corresponds to the boundary between the two 
gapped phases 
in the nonmagnetic ground state, which was obtained 
by the density matrix renormalization group analysis.\cite{rf:15}
According to the present numerical diagonalization approach, 
the boundary is almost independent of the magnetization $m$ 
even in the massless phase. 
In ref.15, the two massive phases were called the singlet 
and Haldane phases, 
which correspond to the gapless1 and gapless2 phases in the magnetic 
state, respectively.
Schematic ground states in the classical limit of gapless1 and gapless2 are
shown in Figs. 4 and 5.
The present analysis suggests that the plateau phase at $m=1/2$ can appear  
only in the magnetization process from the singlet (not Haldane) phase. 
The two gapless phases can be easily explained in the 
classical limit, where they correspond to two different ordered phases
depending on  
whether the rung or diagonal interactions are dominant. 
\begin{figure}
\begin{center}
\leavevmode
\epsfxsize=9.0cm
\epsfysize=3.0cm
\epsfbox{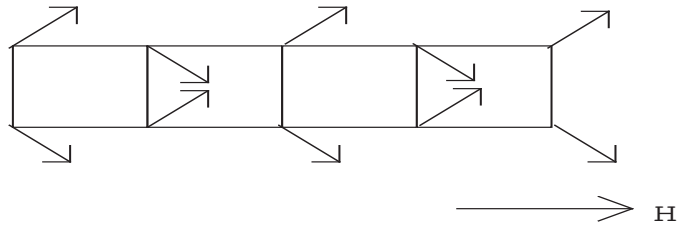}
\caption{Ground state in the classic limit for
0.5$J_{\perp}>{\alpha}J_1$(gapless1)}
\label{fig.4}
\end{center}
\end{figure}
\begin{figure}
\begin{center}
\leavevmode
\epsfxsize=9.0cm
\epsfysize=3.0cm
\epsfbox{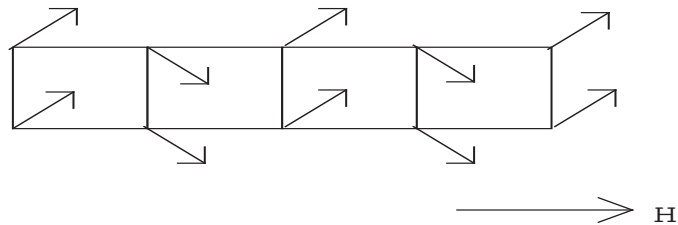}
\caption{Ground state in the classic limit
for 0.5$J_{\perp}<{\alpha}J_1$(gapless2)}
\label{fig.4}
\end{center}
\end{figure}
In the quantum system, instead of each order, 
the characteristic spin correlation function exhibits 
the power-law decay and 
the boundary is slightly deviated from the classical one 
($\alpha J_1=0.5J_{\perp}$) because of the quantum fluctuation. 
In Fig. 3, a necessary condition for
the appearance of the plateau in the present scenario is revealed to 
be that the rung coupling is larger than the leg one ($J_1<J_{\perp}$).  
We derive the magnetization curve in the bulk system 
from the energy $E(L,M)$ of the finite systems. 
In the ground state magnetization process, 
the magnetic field to change $M$ into $M+1$ ($M-1$ into $M$) 
for finite systems is given by 
$H_+(L,M)\equiv E(L,M+1)-E(L,M)$ 
($H_-(L,M)\equiv E(L,M)-E(L,M-1)$). 
We define the thermodynamic limit of $H_{\pm}(L,M)$ as $H_{\pm}(m)$. 
When the system is gapless at $m$, 
the magnetic field to realize $m$ in the bulk system is obtained by 
$H(m)=H_+(m)=H_-(m)$ and the finite size correction is proportional 
to $1/L$. 
On the other hand, 
for the gapped state, $H_+(m)$ and $H_-(m)$ are different and 
correspond to both edges of the plateau, respectively 
($H_-(m)<H_+(m)$). 
In Fig. 6, $H_+(L,M)$ and $H_-(L,M)$ for $m$=1/6, 1/4, 1/3, 1/2, 
2/3 and 3/4 are plotted versus $1/L$ for 
the parameters $(J_1,\alpha )=(0.5,0.6)$ in the plateau phase at
$m=1/2$. 
The plot reveals that 
$H_+(L,M)$ and $H_-(L,M)$ converge to the same value $H(m)$ in the 
infinite length limit except for $m=0$ and 1/2. 
In contrast 
it clearly indicates $H_+({1\over 2})\not= H_-({1\over 2})$ 
which implies that the plateau appears at $m=1/2$.  
So we estimate  
$H_{\pm}({1\over 2})$  
by Shanks transformation\cite{rf:16} 
which is useful for an exponentially
converging sequence.  
The transformation applied for a sequence $P_n$ is defined as 
\begin{eqnarray}
P'_n =\frac{P_{n-1}P_{n+1}-P_n^2}{P_{n-1}+P_{n+1}-2P_n}. 
\end{eqnarray}
Using it, 
we also estimate $H_+(0)$ (the critical field such that the first spin gap 
vanishes).   
Finally, we present the magnetization curve obtained by the
above size scaling method detailed in ref.10,
for $(J_1,\alpha)=$(0.5, 0.6) and (0.5, 0.8) in the plateau phase 
in Fig. 7. 
Both curves exhibit a distinct plateau at $m=1/2$ and
its width increases with increasing $\alpha$, 
as well as the plateau at $m=0$.
\begin{figure}
\begin{center}
\leavevmode
\epsfxsize=7.0cm
\epsfbox{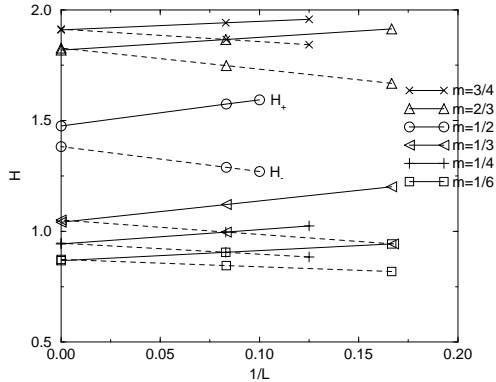}
\caption{$H_+(L,M)$ and $H_-(L,M)$ versus $1/L$}
\label{fig.3}
\end{center}
\end{figure}
\begin{figure}
\begin{center}
\leavevmode
\epsfxsize=7.0cm
\epsfbox{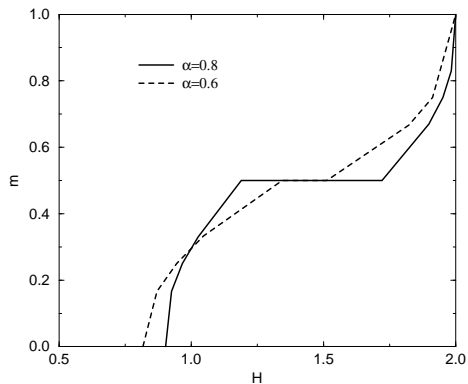}
\caption{Magnetization curves}
\label{fig.4}
\end{center}
\end{figure}
The plateau at $m=1/2$ in the present system corresponds to 
the case when $Q=2$, $S=1$ and $m=1/2$ are held in the condition 
for quantization of the magnetization (\ref{quantization}). 
This implies that the translational symmetry is spontaneously broken 
and the double periodicity of the original unit cell is realized 
in the ground state. 
It is expected that the frustration due to the next-nearest neighbor 
exchange interaction stabilizes the spin structure where the singlet and 
triplet bonds along the rung are alternately placed in the plateau 
phase shown in Fig. 8,
\begin{figure}
\begin{center}
\leavevmode
\epsfxsize=7.0cm
\epsfbox{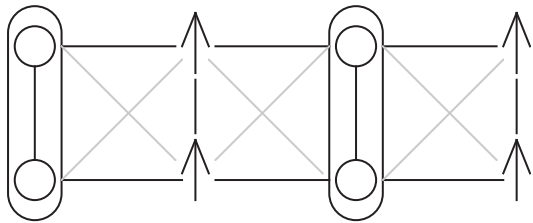}
\caption{Ground state in the plateau at $m=1/2$, 
singlet and triplet bonds alternate along the rung}
\label{fig.5}
\end{center}
\end{figure}
as was discussed for the bond-alternating chain with the 
next-nearest neighbor exchange interaction.\cite{rf:5}
If such a site is realized, the two energy levels with the momentum 
$k=0$ and $\pi$ should be degenerated.
The gap between the two levels for $m=1/2$ is given by 
\begin{eqnarray}
{\Delta}_{\pi}=E_\pi(L,L/2)-E(L,L/2).
\end{eqnarray}
The scaled gap $L{\Delta}_{\pi}$ for $J_{\perp}=1$ and $J_1=0.4$
(the same parameter as Fig. 2) is plotted versus $\alpha$ in Fig. 9.
${\Delta}_{\pi}$ is revealed to decay faster than $1/L$ with increasing
L, which indicates the degeneracy of the two levels in the thermodynamic
limit.
It justifies realization of the ground state specified by Fig. 8 at $m=1/2$.
\begin{figure}
\begin{center}
\leavevmode
\epsfxsize=7.0cm
\epsfbox{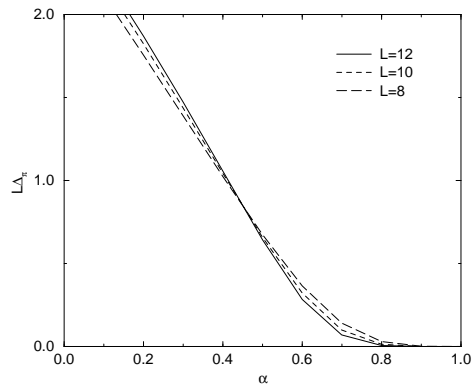}
\caption{Energy gap between the two levels with $k=0$ and $\pi$ at
$m=1/2$}
\label{fig.4}
\end{center}
\end{figure}
The plateau with $Q=2$ was also predicted\cite{rf:13} for the spin ladder 
with the four spin exchange interaction\cite{rf:17}, but the mechanism 
would be different from the present case. 

In summary, 
the exact diagonalization and the size scaling analyses based on CFT 
suggest that the plateau appears at $m=1/2$ in the magnetization 
curve of the spin ladder, 
if $J_{\perp}>J_1$ is satisfied and the next-nearest neighbor 
exchange interaction is sufficiently large (about half the leg coupling). 
We anticipate the observation of the plateau 
in some real materials. 
The magnetization measurement of a rung-dominant spin ladder 
such as Cu$_2$(C$_5$H$_{12}$N$_2$)$_2$Cl$_4$ (ref.18) 
at a high pressure, which might enhance 
the next-nearest neighbor exchange interaction, would also be interesting. 

We wish to thank Drs. K. Okamoto, K. Totsuka, and Prof. K. Nomura 
for fruitful discussions. 
This research was supported in part by a Grant-in-Aid
for the Scientific Research Fund from the Ministry
of Education, Science, Sports and Culture (11440103).
The numerical computation was performed using the facility of the
Supercomputer Center, Institute for Solid State Physics, University of
Tokyo, and the TITPACK ver 2.0 by H. Nishimori.

\end{document}